\documentclass[%
superscriptaddress,
amsmath,
amssymb,
aps,
prb,
twocolumn,
]{revtex4-2}

\usepackage[T1]{fontenc}
\usepackage[utf8]{inputenc}
\usepackage[english]{babel}
\providecommand{\selectlanguage}[1]{}
\usepackage{tabularx,booktabs}
\usepackage[x11names]{xcolor}
\usepackage{graphicx}
\usepackage{dcolumn}
\usepackage{bm}
\usepackage{hyperref}
\hypersetup{
  colorlinks,
  citecolor=Red3,
  linkcolor=black,
  urlcolor=Red3}

\begin{document}


\title{Bright and pure single-photon source in a silicon chip by nanoscale positioning of a color center in a microcavity}

\author{Baptiste Lefaucher}
\altaffiliation{Contributed equally to this work}
\affiliation{Univ. Grenoble Alpes, CEA, Grenoble INP, IRIG, PHELIQS, 38000 Grenoble, France}

\author{Yoann Baron}
\altaffiliation{Contributed equally to this work}
\affiliation{Univ. Grenoble Alpes, CEA-LETI, Grenoble 38000, France}

\author{Jean-Baptiste Jager}
\affiliation{Univ. Grenoble Alpes, CEA, Grenoble INP, IRIG, PHELIQS, 38000 Grenoble, France}

\author{Vincent Calvo}
\affiliation{Univ. Grenoble Alpes, CEA, Grenoble INP, IRIG, PHELIQS, 38000 Grenoble, France}

\author{Christian Elsässer}
\affiliation{Univ. Grenoble Alpes, CEA, Grenoble INP, IRIG, PHELIQS, 38000 Grenoble, France}

\author{Giuliano Coppola}
\affiliation{Univ. Grenoble Alpes, CEA, Grenoble INP, IRIG, PHELIQS, 38000 Grenoble, France}

\author{Frédéric Mazen}
\affiliation{Univ. Grenoble Alpes, CEA-LETI, Grenoble 38000, France}

\author{Sébastien Kerdilès}
\affiliation{Univ. Grenoble Alpes, CEA-LETI, Grenoble 38000, France}

\author{Félix Cache}
\affiliation{Laboratoire Charles Coulomb, Université de Montpellier and CNRS, Montpellier, France}

\author{Anaïs Dréau}
\affiliation{Laboratoire Charles Coulomb, Université de Montpellier and CNRS, Montpellier, France}

\author{Jean-Michel Gérard}
\email{jean-michel.gerard@cea.fr}
\affiliation{Univ. Grenoble Alpes, CEA, Grenoble INP, IRIG, PHELIQS, 38000 Grenoble, France}

\begin{abstract}
We present an all-silicon source of near-infrared linearly-polarized single photons, fabricated by nanoscale positioning of a color center in a silicon-on-insulator microcavity. The color center consists of a single W center, created at a well-defined position by Si$^{+}$ ion implantation through a 150 nm-diameter nanohole in a mask. A circular Bragg grating cavity resonant with the W's zero-phonon line at 1217 nm is fabricated at the same location as the nanohole. By Purcell enhancement of zero-phonon emission, we obtain a photon count rate of $1.29 \pm 0.01$ Mcounts/s at saturation under above-gap continuous-wave excitation with a Debye-Waller factor of $98.6\pm1.4 \%$. A clean photon antibunching behavior is observed up to pump powers ensuring saturation of the W's emission ($g^{(2)}(0)=0.06\pm0.02$ at $P=9.2P_{sat}$), evidencing that the density of additional parasitic fluorescent defects is very low. We also demonstrate the triggered emission of single photons with $93\pm2 \%$  purity under weak pulsed laser excitation. At high pulsed laser power, we reveal a detrimental effect of repumping processes, that could be mitigated using selective pumping schemes in the future. These results represent a major step towards on-demand sources of indistinguishable near-infrared single photons within silicon photonics chips.
\end{abstract}


\maketitle


\section*{Introduction}

\begin{figure*}[hbt!]
  \centering
  \includegraphics[width=470pt]{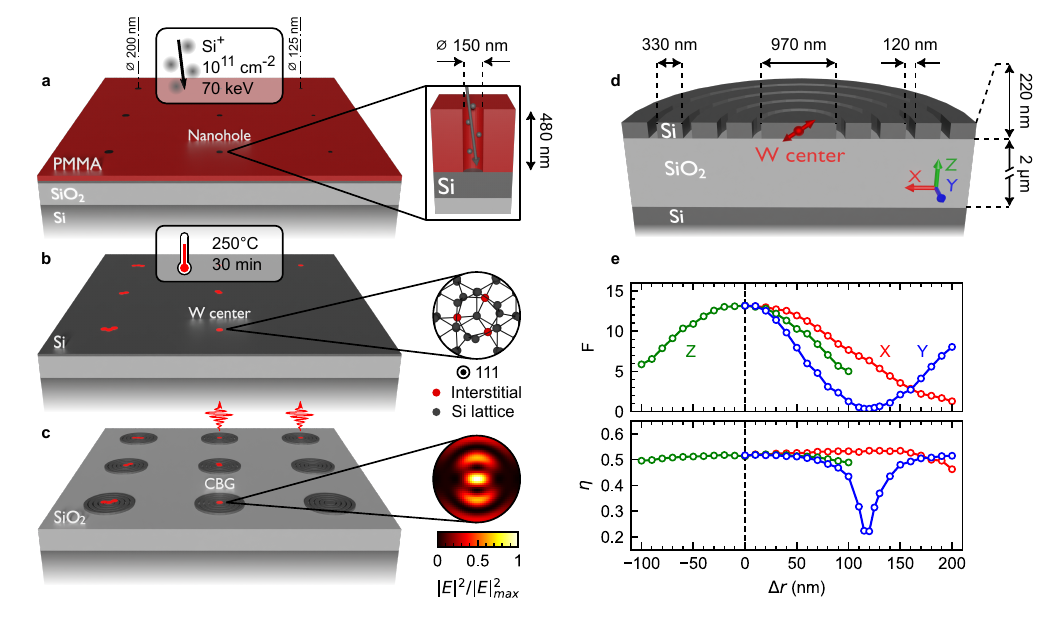}
  \caption{\textbf{Integration of W centers in circular Bragg grating cavities, and expected cavity effects.} 
  \textbf{a,} Implantation of Si$^+$ ions in SOI through a PMMA layer patterned with nanoholes. 
  \textbf{b,} Activation of the W centers by thermal annealing. 
  The microscopic structure of the W center is shown in the inset. 
  \textbf{c,} Fabrication of CBG cavities centered on the nominal coordinates of the nanoholes. 
  The intensity of the electromagnetic field in the central disk is shown in the inset. 
  \textbf{d,} Schematic sectional view of a CBG cavity containing a single W center, with the relevant dimensions. 
  The red double-arrow indicates the orientation of the transition dipole. 
  \textbf{e,} Normalized spontaneous emission rate $F$ and collection efficiency $\eta$ for a numerical aperture of 0.65, as a function of the algebric distance $\Delta r$ between the dipole and the center of the cavity in the coordinate system shown in \textbf{d}.}
  \label{fig:figure_1}
\end{figure*}

\indent The generation, processing, and detection of single photons have enabled pioneering investigations in quantum physics, paving the way to groundbreaking technologies. Silicon quantum photonics, namely the processing of photonic quantum states in a silicon chip, is at the forefront of this revolution \cite{silverstone_silicon_2016, wang_integrated_2020}. However, the potential of silicon photonics for large-scale quantum applications is limited by the difficulty to deterministically generate arbitrary large entangled photon states on-chip. A key requirement to overcome this limitation is to embed sources of indistinguishable single photons operating on demand. While integrated sources of entangled photon pairs based on spontaneous four-wave mixing have enabled major demonstrations \cite{wang_integrated_2020}, these sources suffer from an inherent interplay between efficiency and multi-photon pairs emission. Alternative approaches rely on fiber-coupling of an external source \cite{singh_quantum_2019, bauer_achieving_2021}, and hybrid integration of quantum emitters \cite{dibos_atomic_2018, elshaari_hybrid_2020, katsumi_unidirectional_2021}, albeit with limited efficiency due to coupling losses.

\indent In this context, near-infrared color centers in silicon have emerged as a promising path towards deterministic single-photon sources (SPS) circumventing coupling losses. Due to their emission wavelength in the telecom bands, these point defects also appear as potential sources for chip-based quantum communications. Since the observation of photon antibunching for single color centers in 2020 \cite{redjem_single_2020, hollenbach_engineering_2020}, tremendous efforts have been devoted to optimizing their elaboration processes \cite{baron_single_2022,hollenbach_wafer-scale_2022,zhiyenbayev_scalable_2023, andrini_activation_2024, aberl_all-epitaxial_2024}, and exploring their photophysical properties \cite{durand_broad_2021,baron_detection_2022,higginbottom_optical_2022,prabhu_individually_2023,durand_hopping_2024}. In order to build a SPS, it is necessary to embed the emitter in a photonic cavity to achieve on-demand emission of single photons in a well-defined optical mode using the Purcell effect \cite{moreau_single-mode_2001}. Several single-mode sources in silicon have been demonstrated, based on a T center \cite{islam_cavity-enhanced_2024,johnston_cavity-coupled_2024}, G center \cite{saggio_cavity-enhanced_2024,kim_bright_2025}, G$^{\star}$ center \cite{redjem_all-silicon_2023,ma_nanoscale_2025}, and erbium dopant \cite{gritsch_purcell_2023}.However, in spite of remarkable improvement of emission properties owing to a high Purcell factor, these sources suffer from significant limitations. The T center and erbium dopant are inherently slow emitters \cite{islam_cavity-enhanced_2024,johnston_cavity-coupled_2024}, the G center has a poor quantum efficiency \cite{lefaucher_cavity-enhanced_2023,durand_genuine_2024-1,saggio_cavity-enhanced_2024}, and the fabrication process of the G$^{\star}$ center is not selective \cite{durand_broad_2021}, resulting in poor purity due to the presence of parasitic defects \cite{redjem_all-silicon_2023,ma_nanoscale_2025}.

\indent In recent work, Purcell enhancement has been demonstrated for an ensemble of W centers in circular Bragg grating (CBG) cavities in a silicon-on-insulator (SOI) film \cite{lefaucher_purcell_2024,veetil_enhanced_2024}. On one hand, these results have positioned the W center as a promising candidate to build a SPS, owing to its high quantum efficiency ($\approx 2/3$) and relatively fast zero-phonon emission ($\tau_{ZPL} \approx 134$ ns) in bulk silicon \cite{lefaucher_purcell_2024}. On the other hand, it confirms that CBG cavities, which have been widely used in other material systems \cite{li_efficient_2015,davanco_circular_2011,wang_-demand_2019,kolatschek_bright_2021,iff_purcell-enhanced_2021,froch_coupling_2021,holewa_high-throughput_2024}, are also attractive for building SPS on SOI chips. However, due to a small mode volume, the position of the emitter in the cavity must be controlled accurately to ensure optimal exploitation of the Purcell effect and reproducible fabrication of the SPS.

\indent Here, we present a SPS fabricated by nanoscale positioning of a single W center in a SOI CBG cavity. An array of W centers is created by Si$^+$ ion implantation through a mask patterned with nanoholes \cite{hollenbach_wafer-scale_2022}, then an array of CBG cavities centered at the same location as the nanoholes and resonant with the W zero-phonon line (ZPL) is fabricated. The optical properties of a cavity-coupled W center are studied by photoluminescence (PL) spectroscopy and photon correlation measurements under above-gap continuous-wave (CW) and pulsed excitation. The purity of the source is analyzed as a function of the laser excitation power.

\section*{Sample design and fabrication}

\indent The W centers are created by a well-known combination of Si$^+$ implantation and thermal annealing \cite{yang_sub-bandgap_2010,buckley_optimization_2020}. Similarly to previous work \cite{hollenbach_wafer-scale_2022}, the implantation is performed through a mask patterned with nanoholes whose size is tuned to locally create single W centers. A 480 nm-thick mask of polymethyl methacrylate (PMMA) is deposited onto a commercial piece of SOI wafer, arrays of nanoholes defined by electron beam lithography are etched into the mask, and the sample is broad-beam implanted with Si$^+$ ions. The PMMA mask is then removed by stripping with acetone and the sample is annealed at 200 °C for 30 minutes in N$_2$ atmosphere to activate the emitters (Figure \ref{fig:figure_1}(b)), whose microscopic structure - consisting of three interstitial Si atoms \cite{baron_detection_2022} - is shown in the inset. For a nanohole diameter of 150 nm, single W centers are observed with a yield of 40\% with low background emission ($g{^2}(0) \approx 0.1$) and average count rate of $3.2 \pm 0.8$ kcps. A characterization of the emitters in unpatterned SOI is presented in Section 3 of the Supporting Information (SI).

\indent CBG cavities, consisting of a disk surrounded by four grating rings, are fabricated by electron beam lithography and reactive ion etching at the same location as the nanoholes (Figure \ref{fig:figure_1}(c)). After fabrication, the sample is inspected by scanning electron microscopy to confirm successful fabrication according to design parameters (see Figure S9 in the SI). A map of the electromagnetic field in the central disk of the cavity is shown in the inset. Note that this method differs from the more common strategy of determining the position of preselected emitters by PL mapping, that is more demanding to implement and where the positionning accuracy is limited by optical resolution \cite{ma_nanoscale_2025}.

\indent A sectional view of a CBG cavity containing a single W center is schematically represented in Figure \ref{fig:figure_1}(d), with the relevant dimensions. The red double-arrow represents the orientation of the transition dipole along the $\langle 111 \rangle$ direction \cite{baron_detection_2022}. It is well-known for CBG cavities that the Purcell enhancement factor $F$, defined as the ratio between the emission rate in the cavity and bulk semiconductor, is heavily dependent on the position of the emitter \cite{sapienza_nanoscale_2015}. This sensivity was investigated for our emitter-cavity system using numerical finite-difference time-domain simulations (FDTD, see Section 4 in the SI). Figure \ref{fig:figure_1}(e) shows the Purcell enhancement factor for the ZPL as a function of the distance $\Delta r$ between the center of the cavity and the position of the dipole, in the coordinate system defined in Figure \ref{fig:figure_1}(d). A maximum enhancement factor of 13 is obtained at the center of the cavity. $F$ decreases rapidly against $\Delta r$ and reaches 0.3 for $|\Delta Y| = 120$ nm, emphasizing the requirement for high positioning accuracy. Considering the simulated lateral spread of the ions implanted through a nanohole of 150 nm diameter, we estimate that the W centers have near-unity probability to be located less than 120 nm away from the center of the cavity, i.e. in the vicinity of the main antinode of the electromagnetic field where Purcell enhancement is maximal (see Section 2 in the SI).

\indent The collection efficiency for a cavity-coupled W center was also investigated using FDTD, considering the configuration of our measurement setup. The fraction $\eta$ of dipole emission radiated through a numerical aperture of 0.65 at normal incidence is plotted against $\Delta r$ in Figure \ref{fig:figure_1}(e). A collection efficiency of 51$\%$ is obtained at the center of the cavity. Note that this value could be increased to nearly $100\%$ with the addition of a high-reflectivity mirror below the cavity \cite{rickert_optimized_2019}. The graph also reveals the robustness of the collection efficiency against spatial detuning (with $\eta > 44\%$ for $\Delta r \leq 100$ nm for each three coordinates), except for the "pit" around $|\Delta Y|$ = 120 nm where the emitter is at a node of the electromagnetic field. This feature makes CBG cavities an attractive microstructure to easily increase the photon detection rate from quantum emitters in SOI. Similarly, the collection efficiency is also found robust against spectral detuning (see Figure S7 in the SI).

\section*{Experimental results}

\begin{figure}[!ht]
  \centering
  \includegraphics[width=240pt]{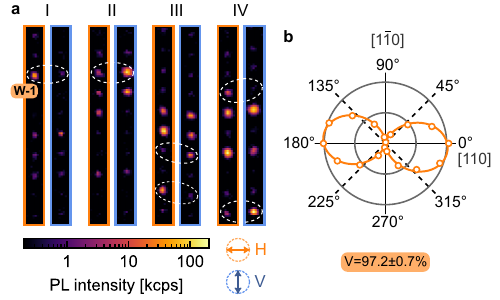}
  \caption{\textbf{Polarization analysis of cavities containing W centers.}
  \textbf{a,} Polarization-selective PL map sets of four arrays of cavities. The maps are recorded at a temperature of 10 K under above-gap CW excitation with a power of 13 $\mu$W.
  \textbf{b,} Polarisation diagram of the spot labeled as 'W-1'.}
  \label{fig:figure_single_W_hunting}
\end{figure}

\indent After fabrication, the sample is characterized by mapping the PL intensity in the emission range of W centers (1200-1300 nm). In the following, we focus on cavities fabricated over nanohole arrays of 150 nm diameter, where single W centers have been identified after the annealing step. On PL maps, the cavities populated by W centers appear as bright spots on a dark background. The number of W centers formed by localized ion implantation being a random variable, finding cavities containing a single center can be a tedious task. Here, we use polarization-selective PL mapping to pre-select potential cavities. As the emission of the W center at normal incidence is linearly polarized along either the [110] or [1$\overline{1}$0] crystal axes of silicon, corresponding to horizontal and vertical polarization in free space in our experimental configuration, single W centers appear as H- or V-polarized spots in PL maps. By recording H- and V-polarized PL on separate maps, potential cavities are identified by polarization contrast. Four pairs of cross-polarized PL maps are displayed in Figure \ref{fig:figure_single_W_hunting}(a). In each pair of maps, one or several bright spots exhibit a clear polarization contrast. By measuring the photon autocorrelation function $g^{(2)}$ in the Hanbury Brown and Twiss configuration, the presence of a single W center in a given spot is revealed by the observation of strong photon antibunching ($g^{(2)}(0)$<0.5). In practice, antibunching is systematically observed for spots with more than 90\% contrast, which is the case for 6 of the 40 spots displayed in the figure (indicated by white ellipses). Thus, we estimate that the end-to-end efficiency of the fabrication process is of the order of $15 \pm 6 \%$.

\indent In the following, we will focus on the spot labeled as W-1 in the map set I, whose polarization diagram (Figure \ref{fig:figure_single_W_hunting}(b)) evidences linear polarization with a contrast of $97.2 \pm 0.7 \%$.

\begin{figure}[!ht]
  \centering
  \includegraphics[width=240pt]{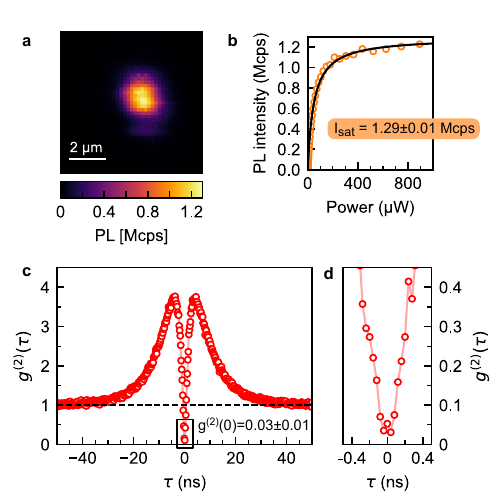}
  \caption{\textbf{Observation of a single W center in a CBG cavity.} 
  \textbf{a,} PL map of a single W center in a CBG cavity at saturation under CW excitation.
  \textbf{b,} PL saturation curve for the W in the CBG cavity under above-gap CW excitation. The PL intensity is integrated over the 1200 - 1300 nm spectral range. The black line is a fit to the saturation function for a two-level system, $I=I_{sat} / ( 1 + P_{sat}/P)$, with $I_{sat}=1.29 \pm 0.01$ Mcps and $P_{sat}=49 \pm 1$ $\mu$W. 
  \textbf{c,} Uncorrected second-order photon correlation histogram of the PL emission for $P = 135$ $\mu$W = 2.8 $P_{sat}$. 
  \textbf{d,} A zoom at short time delays reveals a $g^{(2)}(0)$ value of $0.03 \pm 0.01$}
  \label{fig:figure_2}
\end{figure}

\indent The single-photon emission properties of the W-1 spot under above-gap CW excitation are displayed in Figure \ref{fig:figure_2} (see Section 1 in the SI for details about the measurement configuration). A high-resolution PL map under strong pumping reveals a maximum intensity exceeding 1.2 Mcps when the PL emission is optimally collected (Figure \ref{fig:figure_2}(a)). The power characteristic of the emitter evidences a clear saturation behavior, with a count rate of $1.29 \pm 0.01$ Mcps at saturation. This value, which is 400 times larger than the average count rate obtained for single W centers in planar SOI, is remarkably high compared to previously published results for other Si color centers (See Table \ref{tab:comparison_previous_work} for a comparison). We will show in the next paragraph that this increase results from Purcell enhancement of zero-photon emission and high collection efficiency. To confirm that the PL is emitted by a single W center, the photon autocorrelation function $g^{(2)}$ is measured. A clean antibunching dip is observed without any correction for detectors dark counts or background emission, with $g^{(2)}(0)=0.03\pm0.01$ at $P=2.8P_{sat}$ (Figure \ref{fig:figure_2}(c, d)) and $g^{(2)}(0)=0.06\pm0.02$ at $P=9.2P_{sat}$ (Figure S10 in the SI), evidencing high-purity single-photon emission up to pumping powers ensuring saturation of the emitter. These $g^{(2)}(0)$ values are among the best reported values for Si color centers, as shown in Table \ref{tab:comparison_previous_work}, indicating that the source is nearly free of additional fluorescent defects. The autocorrelation histogram also exhibits a bunching behavior ($g^{(2)}(\tau) > 1$) at non-zero delay. This phenomenon, already reported for single W centers \cite{baron_detection_2022, hollenbach_wafer-scale_2022}, is attributed to a metastable state coupled to the excited state of the emitter. It is also worth mentionning that, as the measurements are performed under strong pumping, the width of the antibunching dip is defined by the average repumping time and does not correspond to the lifetime of the excited state \cite{baron_detection_2022}.

\begin{table}[!t]
  \caption{Comparison between this work and previous reports of cavity-enhanced single-photon emitters in silicon under CW pumping. 
  Em.: emitter. 
  Cav.: cavity. 
  Pos.: nanoscale positioning. 
  PhC: photonic crystal cavity.
  We give for T and W results obtained at 2 different powers that either provide maximum $I_{PL}$ or minimum $g^{(2)}(0)$ value.}
  \label{tab:comparison_previous_work}
  \begin{ruledtabular}
  \begin{tabularx}{\columnwidth}{c c c c c c c}
  Em. & Cav. & $I_{PL}$ (kcps) & $\tau$ (ns) & $g^{(2)}(0)$ & Pos. & Ref. \\
  \hline
  T           & 1D PhC  & 20                & 169               & 0.07          & No    & \cite{islam_cavity-enhanced_2024} \\
              &         & 236               &                   & 0.35          &       &  \\
  \hline
  G           & 2D PhC  & 20                & 6                 & 0.03          & No    & \cite{saggio_cavity-enhanced_2024} \\
  \hline
  G$^{\star}$ & 2D PhC & 20 & 6.7               & 0.30          & No    & \cite{redjem_all-silicon_2023} \\
  \hline
  G$^{\star}$ & CBG  & 260 & 4.3 & 0.37 & Cav. & \cite{ma_nanoscale_2025} \\
  \hline
  W           & CBG     & 940               & 7                 & 0.03          & Em.+Cav.   & This \\
              &         & 1200              &                   & 0.06          &       & work  \\
  \end{tabularx}
  \end{ruledtabular}
\end{table}

\begin{figure}[!ht]
  \centering
  \includegraphics[width=240pt]{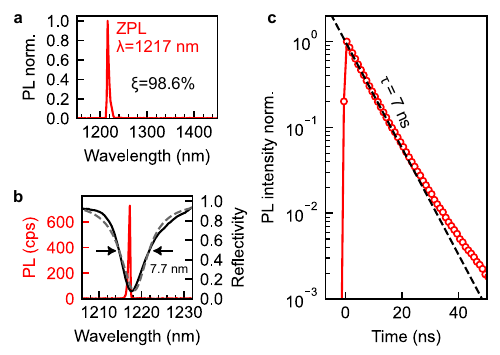}
  \caption{\textbf{Optical properties of a CBG-coupled W center.} 
  \textbf{a,} PL spectrum in the 1150 - 1450 nm spectral range. 
  \textbf{b,} High-resolution PL spectrum and white-light reflectometry measurement of the cavity. 
  The dashed grey line is a fit to a Lorentzian profile. 
  \textbf{c,} PL decay of the emitter under triggered excitation with a repetition rate of 5 MHz and 17.5 $\mu$W average power.}
  \label{fig:figure_3}
\end{figure}

\indent To understand cavity effects quantitatively, we investigated the optical properties of the W-1 center and the spectral resonance of the emitter-cavity system. The PL spectrum of the emitter over the 1140-1450 nm spectral range reveals that the ZPL accounts for $\xi = 98.6\pm 1.4\%$ of the PL intensity (Figure \ref{fig:figure_3}(a)). This value should be compared with 39\% for isolated W in unpatterned SOI, which unveils selective enhancement of the ZPL by the cavity. To confirm this hypothesis, the modal properties of the cavity have been probed by performing reflection spectroscopy under white light illumination. The reflectivity curve features a dip around 1218 nm fitting to a Lorentzian profile with a full width at half maximum (FWHM) $\Delta \lambda_m = 7.7 \pm 0.2$ nm, corresponding to a quality factor $Q = 158 \pm 5$ (Figure \ref{fig:figure_3}(b)). This value is in good agreement with the quality factor estimate of 167 obtained from FDTD simulations. A comparison between the reflectivity curve and a high-resolution PL spectrum shows that the CBG cavity mode is well-matched to the ZPL of the W-1 center. We conclude that the selective enhancement of the ZPL is induced by (1) efficient coupling of zero-phonon emission into the cavity mode, whose narrow far-field radiation pattern ensures high collection efficiency, and (2) possibly acceleration of zero-phonon emission by Purcell effect.

\indent To estimate the magnitude of the Purcell effect, we study the recombination dynamics of the W-1 center. 
The PL decay of the emitter under pulsed excitation at 17.5 $\mu$W with a 5 MHz repetition rate is displayed in Figure \ref{fig:figure_3}(c). 
The intensity decays nearly mono-exponentially with a characteristic time $\tau = 7 \pm 1$ ns. 
This value should be compared with the characteristic time of 34 ns obtained for ensembles of W centers in unpatterned SOI \cite{buckley_optimization_2020, lefaucher_purcell_2024,veetil_enhanced_2024}. The acceleration of the decay compared to this reference value has two possible origins: Purcell enhancement of zero-phonon emission, and/or activation of extra non-radiative recombination channels due to the interaction between the emitter and its environment. 
This second possibility cannot be neglected \textit{a priori} as a significant lifetime dispersion has been reported for single W centers in 60 nm-thick SOI \cite{baron_detection_2022}, although this dispersion can likely be attributed to surface effects in such a thin SOI layer. 
A detailed analysis of the photon budget taking into account the setup transmission provides a lower-bound estimate of 1/18 ns$^{-1}$ for the zero-phonon emission rate (see Section 5 in the SI). 
By comparison with a previous estimate of 1/134 ns$^{-1}$ in bulk Si \cite{lefaucher_purcell_2024}, we conclude that the zero-phonon emission rate is increased by a factor $F > 7.2$. 
Therefore, the acceleration of the PL decay results predominantly (and possibly entirely) from the Purcell effect. 
Additionally, according to Figure \ref{fig:figure_1}(e), we can also conclude that the W-1 center is positioned less than 100 nm away from the cavity center.

\begin{figure*}[!ht]
  \centering
  \includegraphics[width=6.25in]{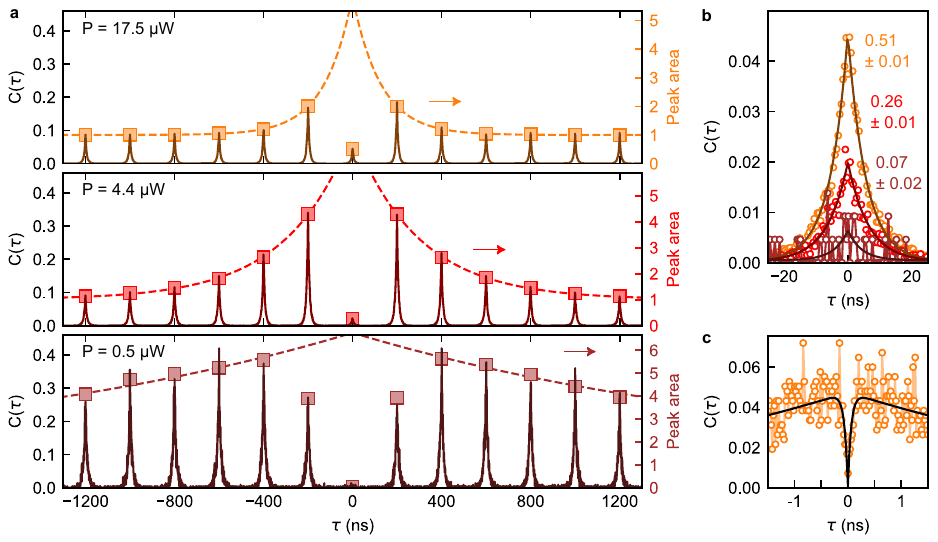}
  \caption{\textbf{Triggered single photons from a CBG-coupled W center.} 
  \textbf{a,} Second-order correlation histogram $C(\tau)$ for $P$ = 0.5, 4.4 and 17.5 $\mu$W. 
  The normalized area of each peak is given by the coordinate of the filled squares on the right axis. 
  The dashed lines represent a fit of the peaks area to Equation \ref{eqn:photoblinking}. 
  \textbf{b,} Zoom on the central peak. The numbers indicate the area of the peak for each power value. 
  \textbf{c,} Zoom on the central peak for $P$ = 17.5 $\mu$W with 20 ps time-bins.}
  \label{fig:figure_4}
\end{figure*}

\indent The bright emission of antibunched photons under CW excitation unveils the strong potential of W centers for building a deterministic SPS in SOI. Along this line, we have explored the triggered emission of single photons under pulsed excitation. Figure \ref{fig:figure_4}(a) shows uncorrected autocorrelation histograms obtained for different pump powers for a 5 MHz repetition rate. The normalized area $A_m$ as a function of time delay $mT$, where $T$ is the laser repetition period, of each peak is given by the coordinate of the squares on the right axis. In the low power regime, the peak at zero-delay has an area $A_m=0.07 \pm 0.02$, demonstrating the triggered emission of single-photon pulses with $93\pm2 \%$ purity. As for the measurement under CW excitation, a bunching behavior is observed ($A_m$>1), indicating photoblinking of the source. Note that this effect is a rather common issue for solid-state quantum emitters and is expected for W, as this color center possesses a metastable excited state \cite{baron_detection_2022}. The area $A_m$ of the peaks follows an exponential law given by:
\begin{equation}
  \label{eqn:photoblinking}
  A_{m \neq 0} = 1 + \frac{\tau_{off}}{\tau_{on}} e^{-(1/\tau_{off} + 1/\tau_{on})|mT|}
\end{equation}
\noindent where $\tau_{on}$ and $\tau_{off}$ are the average times of stay of the emitter in the "bright" and "dark" states, respectively \cite{santori_triggered_2001}. For $P$ = 0.5, 4.4, and 17.5 $\mu$W, $\tau_{on}$ = $2290\pm40$, $341\pm5$, and $153\pm3$ ns, and $\tau_{off}$ = $13200\pm400$, $2200\pm30$, and $741\pm5$ ns, pointing towards an acceleration of the blinking with increasing power. Interestingly, the peaks for $|m|=1$ have a smaller area than expected from the model in the low-power regime. This effect, which has been reported for single InAs QD under above-gap excitation \cite{santori_single-photon_2004}, may be related to a two-step transition from the ground state to the excited state, which is indeed predicted by ab-initio calculation \cite{baron_detection_2022}.

\indent The area $A_0$ of the central peak increases with power, as shown in Figure \ref{fig:figure_4}(b). For $P$ = 17.5 $\mu$W, the area of the peak reaches $0.51 \pm 0.01$, meaning that about two photons are emitted on average per excitation pulse. A zoom on the peak with 20 ps time bins (Figure \ref{fig:figure_4}(c)) reveals an antibunching dip with a characteristic time of $\approx$ 60 ps (that disappears in Figure \ref{fig:figure_4}(b) due to rebinning). This antibunching behavior indicates that the two photons are emitted successively. An explanation is that, under strong above-gap pumping condition, a second exciton can be captured after the emission of a first photon, as reported for other quantum emitters \cite{dalgarno_hole_2008}. This phenomenon is not surprising here, due to the optical injection of a large population of carriers by the pump pulse in the central disk of the cavity. As for QD-SPS, we expect to overcome this limitation through the implementation of resonant or quasi-resonant pumping schemes.

\section*{Present limitations and development prospects}

In spite of promising results under CW excitation, the generation of single photon pulses remains relatively inefficient. At a repetition rate of 5 MHz, the PL count rate reaches 52 kcps at saturation, corresponding to a first-lens efficiency of a few percents (see Section 5 of the SI). The main factors limiting this value at this stage are easily identified:
(1) the emitter is optically active only 17\% of the time, which reduces its average efficiency by a factor of almost 6; 
(2) our CBG cavity funnels only half of the emission towards our collection optics; 
(3) non-radiative recombination limits the emitter's quantum efficiency. For a single W center ideally placed at the center of a CBG cavity, the probability to generate a ZPL photon upon excitation is expected to reach 90\% at most \cite{lefaucher_purcell_2024}.

\indent While point (1) and the issue of blinking has already been discussed, we see that points (2) and (3) can be radically improved by replacing our low-Fp CBG cavity by a high-Fp waveguide-coupled photonic crystal cavity, such as a nanobeam cavity \cite{velha_2007,islam_cavity-enhanced_2024,johnston_cavity-coupled_2024}. As the mode volume is similar for PhC and CBG cavities, the nanoscale positioning method is expected to apply equally well for both types of cavity. It should also be noted that the accuracy of the positioning can be further improved by increasing the Si$^+$ dose and decreasing the holes diameter. Proper engineering of light extraction towards the waveguide mode and a large but realistic Purcell factor of 150 would bring the zero-phonon emission rate above 10$^{9}$ s$^{-1}$. This would ensure that the quantum efficiency of the ZPL, the spontaneous emission factor $\beta$ and the efficiency of the waveguide-coupled SPS are all near unity. Single photon coherence and indistinguishability are also required properties for applications to quantum simulations and computing. Until now, Hong-Ou-Mandel interference have only been demonstrated over relatively narrow time-windows, much shorter than the emitter lifetime \cite{komza_indistinguishable_2024}. Combining such a large Purcell acceleration of zero-phonon emission, with strategies aiming at reducing dephasing processes (such as quasi-resonant pumping or the application of a static electric field), will likely permit to generate highly-indistinguishable photons on demand.

\indent From a technological point of view, our strategy is compatible with the integration of multiple single-photon sources connected to the same photonic circuit for quantum applications. In a photonic chip comprising all processing elements and single-photon detectors, W centers spots could be created at defined positions in a dedicated area of the chip. Single W centers would then be mapped and connected to the circuit by fabrication of waveguide-coupled PhC cavities. In order to fully benefit from their large Purcell-enhancement factor, one has to implement means enabling a fine control of the spectral detuning between W's ZPL and the high-Q cavity mode. In this context, electrical and strain tuning of the ZPL have been recently demonstrated for G centers \cite{day_electrical_2024, ristori_strain_2024}.

\section*{Conclusion}

\indent In conclusion, we have demonstrated a source of near-infrared linearly-polarized single photons obtained by nanoscale positioning of a W center in a SOI CBG cavity. Si$^+$ implantation through a mask patterned with nanoholes allows for selective creation of single W centers, positioned in the vicinity of the main antinode of the resonant mode. The cavity-coupled emitter has a PL intensity at saturation of $1.29 \pm 0.01$ Mcps together, with a $g^{(2)}(0)$ value inferior to 0.06 under above-gap CW excitation up to $P=9.2P_{sat}$, and with a ZPL fraction of $98.6\pm1.4 \%$. Under weak pumping condition, the triggered emission of single-photon pulses is demonstrated with $93\pm2 \%$ purity. These results highlight the great potential of W centers for silicon quantum photonics. They pave the way towards the scalable integration of bright sources of indistinguishable single photons on integrated photonic chips, based on single W centers deterministically positioned in high-$F_p$ waveguide-coupled photonic crystal nanocavities.

\section*{Methods}

\noindent \textbf{Sample preparation.}
The silicon layer and buried oxide layer of our SOI have a thickness of 220 nm and 2 $\mu$m, respectively. 
The PMMA layer is chosen thick enough (480 nm) to stop the Si$^+$ ions with nearly 100$\%$ probability during the implantation (see Figure S3 in the Supporting Information). 
The nanoholes etched into the mask have diameters ranging from 125 nm to 2 $\mu$m (see Table S1 in the Supporting Information), and 10 $\mu$m pitch between the holes. 
The implantation of Si$^+$ ions is performed at an energy of 70 keV, with a dose of 10$^{11}$ cm$^{-2}$ and 7° tilt angle with respect to the normal of the sample to avoid channeling.
\newline
\newline
\noindent \textbf{Experimental setup.}
The full optical measurement setup is schematically represented in SI Section 1.
The sample is cooled at 10 K in a helium-flow cryostat (Oxford Instruments Microstat HiRes). 
Continuous-wave and pulsed optical excitation are provided by a semiconductor laser (PicoQuant LDH-IB-485-B) operating at 485 nm. 
In pulsed operation mode, the pulses have a full-width at half-maximum (FWHM) of 100 ps and tunable repetition rate. 
The laser beam is coupled into a single-mode fiber for spatial filtering, and collimated using a microscope objective mounted onto a translation stage in order to finely control the divergence of the beam. 
The laser is focused onto the sample to a 1/e$^2$-spot diameter of $\approx$ 700 nm using a microscope objective with a numerical aperture of 0.65 (Olympus LCPLN50XIR). 
The objective is mounted onto a tri-axial positioner allowing for 50 nm translation steps for photoluminescence (PL) mapping. 
A visualization line is accessible using a removable beamsplitter, in order to obtain an image of the sample on a CCD camera for alignment procedures.

\indent The PL is collected using the microscope objective and separated from the laser using a dichroic mirror with a cutoff wavelength of 650 nm. 
The 1200 - 1300 nm spectral range is selected through a combination of longpass and shortpass filters, and the PL is focused into a single-mode fiber using a reflective collimator (Thorlabs RC04APC-P01). 
The PL is detected using superconducting nanowire single-photon detectors (SNSPD, ID Quantique ID281), with a detection efficiency of 81$\%$ for Detector 1 at 1310 nm (resp. 86$\%$ for Detector 2), dark count rate of 9 cps (resp. 8 cps), and timing jitter of 29 ps (resp. 30 ps). 
The polarization of the photons impinging on the SNSPD is controlled using a half-waveplate and a fiber polarization controller for optimized detection efficiency for both detectors. 
The photon detection events are counted by a time-correlated single-photon counting card (PicoQuant MultiHarp 150). 
For second-order correlation measurements, the card is operated in a time-tagged time-resolved mode.

\indent For spectral measurements, the PL is dispersed by a grating with 900 lines/mm in a spectrometer (Horiba JY iHR320) and detected by a linear InGaAs array (Horiba 81043 Spectrum One). 
The system has a spectral resolution of 0.4 nm.

\indent For polarization-resolved measurements, a motorized half-waveplate is placed in front of a polarizing beamsplitter cube (PBS). 
The PBS has a transmission of 98.2$\%$ for P-polarized light and 0.135$\%$ for S-polarized light at 1218 nm. 
The intensity is recorded as a function of the angle $\theta$ between the axis of the half-waveplate and the transmission axis of the PBS. 
The intensity is plotted as a function of 2$\theta$ to reconstruct the polarization diagram of the PL.

\indent For reflectometry measurements, white-light from a tungsten source is injected into a single-mode fiber and collimated using a reflective collimator. 
A 50:50 removable beamsplitter cube is used to inject the white beam into the excitation line, and the reflected light is collected using the same collection line as for the PL.

\section*{Data availability}

The data supporting the findings of this study are available from the corresponding author upon reasonable request.

\begin{acknowledgements}
We thank Guillaume Cassabois and Vincent Jacques for fruitful discussions, as well as Krithika V. R. and Alicia Morandini for experimental support.
This work is supported by the Plan France 2030 through the project OQuLuS (No. ANR-22-PETQ-0013), the French National Research Agency (ANR) through the projects OCTOPUS (No. ANR-18-CE47-0013-01) and WOUAH (No. ANR-24-CE47-4667), CEA through the PTC-MP “W-TeQ” internal project, and the European Research Council (ERC) under the European Union's Horizon 2020 research and innovation programme (project SILEQS, Grant No. 101042075). 
B.L. is supported by the “Program QuantForm-UGA No. ANR-21-CMAQ-003 France 2030” and “Laboratoire d'Excellence LANEF No. ANR-10-LABX-51-01”. C.E. is supported by the MSCA Cofund QuanG (Grant Number: 101081458), funded by the European Union. The authors acknowledge the assistance of the staff of the Grenoble Advanced Technological Platform (PTA).
\end{acknowledgements}

\section*{Supplementary Information Available}
See the Supporting Information for additional details about simulation of ion implantation, optical properties of single W centers in unpatterned SOI, finite-difference time-domain simulations, estimation of the quantum efficiency and zero-phonon emission rate of the W-1 center, SEM image, and photon autocorrelation measurement at high power.

\bibliography{bib_file}

\end{document}



\title{Supporting Information:\\Bright and pure single-photon source in a silicon chip by nanoscale positioning of a color center in a microcavity}

\author{Baptiste Lefaucher}
\altaffiliation{Contributed equally to this work}
\affiliation{Univ. Grenoble Alpes, CEA, Grenoble INP, IRIG, PHELIQS, 38000 Grenoble, France}

\author{Yoann Baron}
\altaffiliation{Contributed equally to this work}
\affiliation{Univ. Grenoble Alpes, CEA-LETI, Grenoble 38000, France}

\author{Jean-Baptiste Jager}
\affiliation{Univ. Grenoble Alpes, CEA, Grenoble INP, IRIG, PHELIQS, 38000 Grenoble, France}

\author{Vincent Calvo}
\affiliation{Univ. Grenoble Alpes, CEA, Grenoble INP, IRIG, PHELIQS, 38000 Grenoble, France}

\author{Christian Elsässer}
\affiliation{Univ. Grenoble Alpes, CEA, Grenoble INP, IRIG, PHELIQS, 38000 Grenoble, France}

\author{Giuliano Coppola}
\affiliation{Univ. Grenoble Alpes, CEA, Grenoble INP, IRIG, PHELIQS, 38000 Grenoble, France}

\author{Frédéric Mazen}
\affiliation{Univ. Grenoble Alpes, CEA-LETI, Grenoble 38000, France}

\author{Sébastien Kerdilès}
\affiliation{Univ. Grenoble Alpes, CEA-LETI, Grenoble 38000, France}

\author{Félix Cache}
\affiliation{Laboratoire Charles Coulomb, Université de Montpellier and CNRS, Montpellier, France}

\author{Anaïs Dréau}
\affiliation{Laboratoire Charles Coulomb, Université de Montpellier and CNRS, Montpellier, France}

\author{Jean-Michel Gérard}
\email{jean-michel.gerard@cea.fr}
\affiliation{Univ. Grenoble Alpes, CEA, Grenoble INP, IRIG, PHELIQS, 38000 Grenoble, France}

\maketitle
\tableofcontents
\newpage


\section{Optical measurement setup}

\begin{figure}[ht]
  \centering
  \includegraphics[width=504pt]{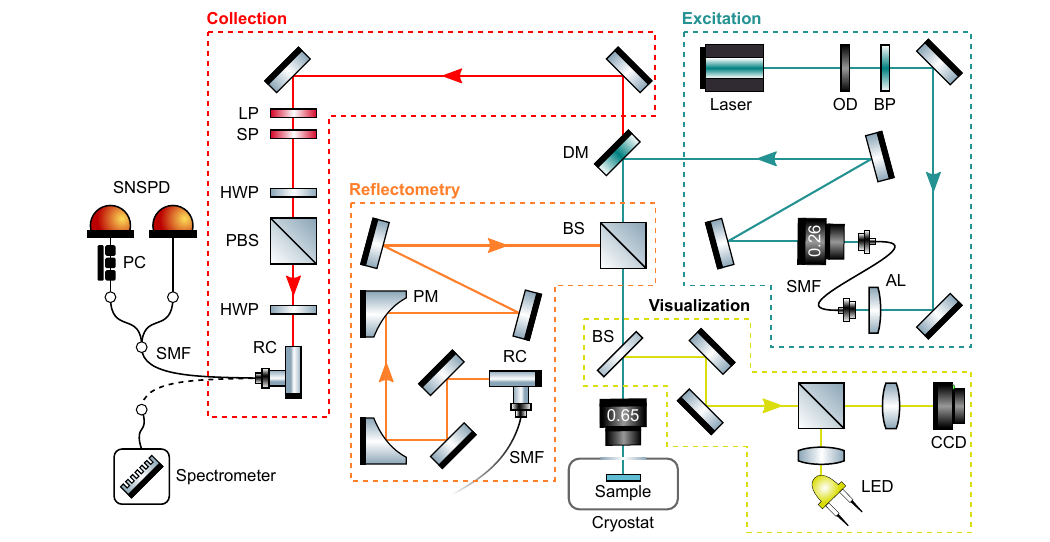}
  \caption{\textbf{Sketch of the optical measurement setup.}
  AL: aspheric lens.
  BP: bandpass filter.
  BS: beamsplitter.
  DM: dichroic mirror.
  HWP: half waveplate.
  LP: longpass filter.
  OD: optical density.
  PBS: polarizing beamsplitter.
  PC: polarization controller.
  PM: parabolic mirror.
  RC: reflective collimator.
  SMF: single mode fiber.
  SNSPD: superconducting nanowire single photon detector.
  SP: shortpass filter.}
  \label{fig:figure_supp_setup}
\end{figure}

\newpage

\section{Simulation of the Si$^+$ ion implantation}

\subsection{Distribution of ions in the silicon-on-insulator layer}

\begin{figure}[ht]
  \centering
  \includegraphics[width=300pt]{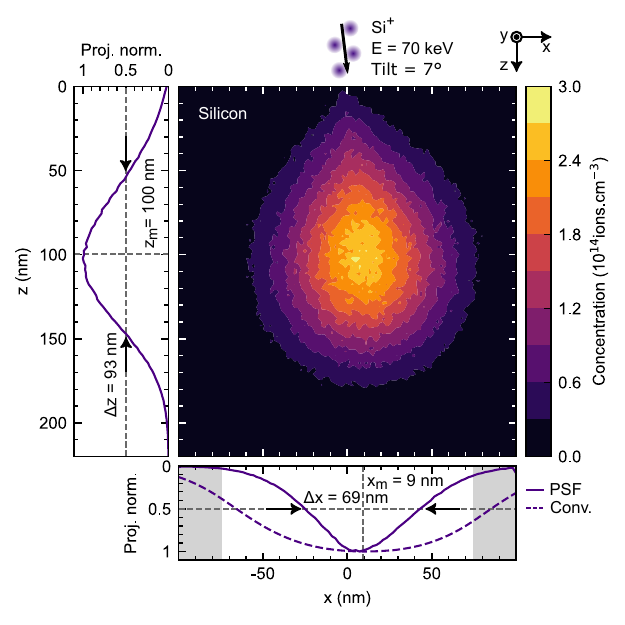}
  \caption{\textbf{Spatial distribution of Si$^+$ ions implanted in the silicon-on-insulator layer.} 
  The ions are implanted with an energy of 70 keV, 7° tilt, and a dose of 10$^{11}$ cm$^{-2}$. 
  The data are obtained from a 3D Monte Carlo simulation with 10$^6$ repetitions using the SRIM software. 
  The left panel displays the projection of the distribution on the $z$ axis. 
  In the bottom panel, the full line indicates the point spread function (PSF) of the implanted ions, i.e. the projection of the distribution on the $x$ axis for a single impact point. 
  The dashed line corresponds to the convolution of the PSF with a disk of 150 nm diameter, delimited by the grey shaded area, corresponding to the nanohole diameter in the PMMA mask.}
  \label{fig:figure_supp_srim}
\end{figure}

\indent A 3D Monte Carlo simulation of the Si$^+$ implantation of the silicon-on-insulator (SOI) wafer is performed using the Stopping and Range of Ions in Matter software (SRIM) \cite{note_srim}. The result of the simulation is presented in Figure \ref{fig:figure_supp_srim}. The point of coordinates ($x,y,z$)=(0,0,0), indicated by the arrow on top, is the impact point of the ions on the top surface of the silicon layer. The ions are implanted with an energy of 70 keV, with a tilt of 7° with respect to the $z$ axis in the $xz$ plane. The simulation is performed with 10$^6$ repetitions.

\indent The 2D plot represents the concentration of ions projected onto the $xz$ plane, for a dose of 10$^{11}$ ions.cm$^{-2}$. The projection of the distribution on the $z$ axis is displayed in the left panel, showing that the ions are implanted at an average depth $z_m$ = 100 nm with a FWHM of 93 nm. The projection on the $x$ axis, in the bottom panel, shows that the ions are implanted at an average distance $x_m$ = 9 nm from the impact point along the $x$ axis, with a FWHM of 69 nm.

\indent In previous work using similar implantation dose and annealing temperature, we have demonstrated that the distribution of W centers matches the vertical distribution of implanted Si$^+$ ions relatively well \cite{lefaucher_purcell_2024}. The lateral spread of the W centers can therefore be tentatively estimated as the convolution between the lateral point spread function of the Si$^+$ ions (plotted in full line in the bottom panel) and a disk of 150 nm diameter, corresponding to the nanohole diameter in the PMMA mask. The resulting distribution is plotted in dashed line. Considering that the cavities are positioned with an accuracy better than 30 nm, the W centers have near-unity probability to be located less than 120 nm away from the center of the cavity, and therefore couple to the main antinode of the electromagnetic field (see Figure 1(e) in the main text). This point is confirmed by the observation of numerous bright W centers in the sample (see Figure 2 in the article).

\subsection{Distribution of ions in the PMMA layer}

\begin{figure}[ht]
  \centering
  \includegraphics[width=8cm]{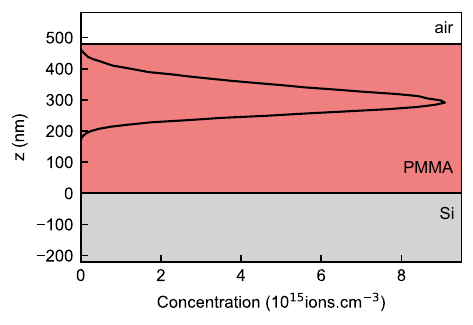}
  \caption{\textbf{Distribution of Si$^+$ ions implanted in the PMMA layer,} obtained using the procedure described in the previous section. 
  The PMMA layer has a thickness of 480 nm. 
  The ions have an average position $z_m$ = 306 nm, corresponding to a projecting range $R_p$ = 174 nm from the air/PMMA interface.}
  \label{fig:figure_supp_srim_pmma}
\end{figure}

\newpage

\section{Single W centers in unpatterned silicon-on-insulator}

\subsection{Diameter of the nanoholes}

\begin{table}[ht!]
  \begin{center}
  \caption{Diameter of the nanoholes in the polymethyl methacrylate (PMMA) mask for Si$^+$ implantation. Each set of nanoholes contains 12+ different diameter values. $n$ is used as a label to identify the diameter of a nanohole as a function of its position in the set.}
  \label{tab:pmma_hole_sizes}
  \begin{tabular}{c c c c c c c c c c c c c c c c}
  \hline
  $n$ & 01 & 02 & 03 & 04 & 05 & 06 & 07 & 08 & 09 & 10 & 11 & 12 \\
  \hline
  $D$ (nm) & 2000 & 1000 & 800 & 600 & 500 & 400 & 350 & 300 & 250 & 200 & 150 & 125 \\
  \hline
  \end{tabular}
  \end{center}
  \end{table}

\subsection{Photoluminescence map before fabrication of the cavities}

\begin{figure}[ht]
  \centering
  \includegraphics[width=300pt]{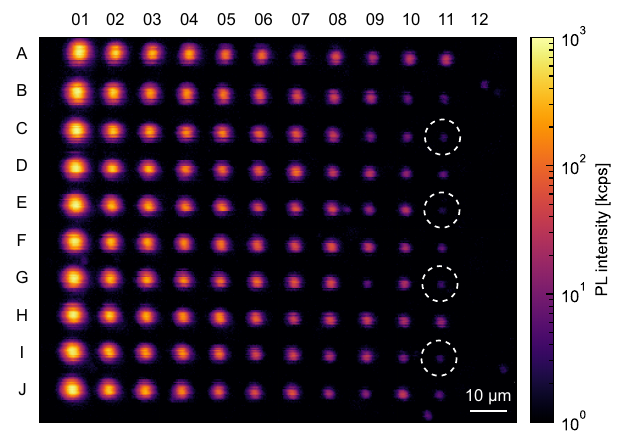}
  \caption{\textbf{PL map of the unpatterned silicon-on-insulator wafer,} at 10 K, under continuous-wave excitation with a power of 135 $\mu$W. 
  The integer labels on top of the map refer to the diameter of the nanoholes for each column (see Table \ref{tab:pmma_hole_sizes}). 
  The letters on the left side are used as labels for the lines, all corresponding to the same set of nanoholes. 
  The dashed circles in the 11$^{th}$ column ($D$ = 150 nm) indicate PL spots that have been identified as single W centers. 
  These spots have an average background-corrected intensity of 3.2 $\pm$ 0.8 kcps.}
  \label{fig:figure_supp_scan_unpatterned}
\end{figure}

\newpage

\subsection{Optical properties of a single W center}

\begin{figure}[ht]
  \centering
  \includegraphics[width=240pt]{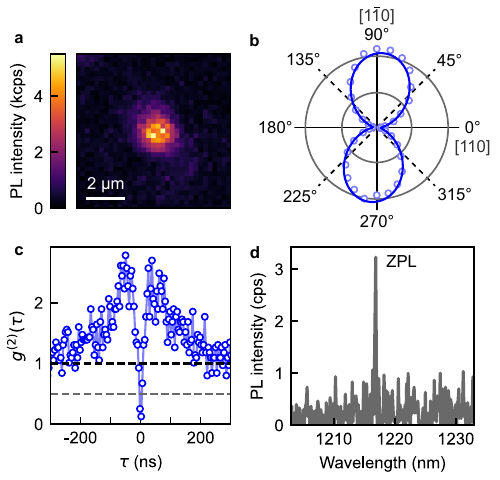}
  \caption{\textbf{Optical properties of a single W center in unpatterned silicon-on-insulator.} 
  \textbf{a,} PL map. 
  \textbf{b,} Polarization diagram of the PL. Here, the PL is polarized along the [1$\overline{1}0$] axis. 
  For all studied W centers, the polarization is oriented along [110] and [1$\overline{1}0$], in agreement with previous work \cite{baron_detection_2022}.
  \textbf{c,} Second-order correlation histogram with 4 ns time bins. 
  \textbf{d,} PL spectrum. The sharp peak at 1217 nm is unambiguously identified as the zero-phonon line (ZPL) of a single W center. 
  The map, polarization diagram and spectrum are obtained under continuous-wave excitation with a power of 135 $\mu$W. 
  The second-order correlation histogram is obtained with a power of 13 $\mu$W. 
  All measurements are performed at 10 K.}
  \label{fig:figure_supp_properties_unpatterned}
\end{figure}

\newpage
\section{Finite-Difference Time-Domain simulations}

\subsection{Optimization of the circular Bragg grating cavities}

\begin{figure}[ht]
  \centering
  \includegraphics[width=504pt]{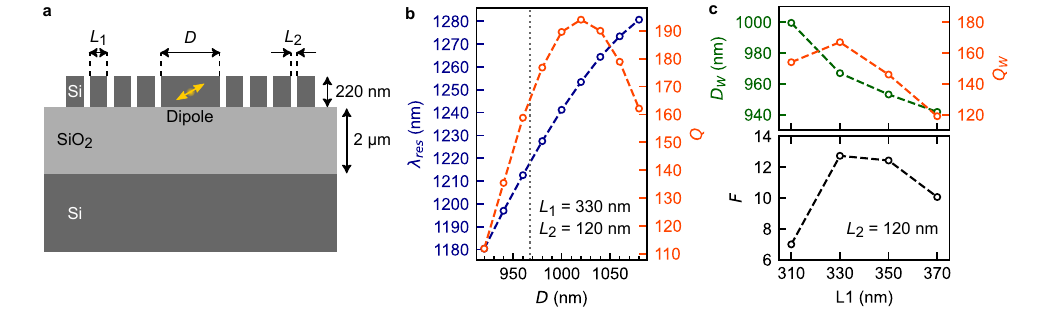}
  \caption{\textbf{Optimization of the CBG cavities.} 
  \textbf{a,} Schematic of the design with the relevant dimensions and parameters. 
  \textbf{b,} Wavelength $\lambda_{res}$ and quality factor $Q$ of the resonant mode as a function of the diameter $D$ of the central disk, for $L_1$ = 330 nm and $L_2$ = 120 nm. 
  \textbf{c,} Diameter $D_W$ allowing resonance at 1218 nm as a function of $L_1$, and corresponding quality factor $Q_W$. For each set of parameters, the Purcell factor $F$ is calculated for a W center located at the center of the cavity.}
  \label{fig:figure_supp_optimization}
\end{figure}

\indent Finite-Difference Time-Domain (FDTD) simulations are performed using the FULLWAVE solver from Synopsys in order to optimize the Circular Bragg Grating (CBG) cavity design. The simulations consist in calculation of transient and steady-state electromagnetic fields by resolving time-dependent Maxwell equations in a discretized volume of space. A sectional view of the 3D simulation box is represented in Figure \ref{fig:figure_supp_optimization}(a). The top silicon layer has a fixed thickness of 220 nm, and the CBG cavity has a free diameter $D$, rings width $L_1$ and gap-width $L_2$. The buried oxide layer has a thickness of 2 $\mu$m. A dipole emitting at 1218 nm, oriented along the $\langle 111 \rangle$ direction, is placed at the core of the cavity.

\indent The wavelength $\lambda_{res}$ and quality factor $Q$ of the resonant mode are determined using a cavity ringdown calculation. At time $t=0$, the dipole flashes a pulse of light. A monitor placed in the vicinity of the dipole records the amplitude of the electric field as a function of time, until the field has almost entirely escaped from the cavity. The spectral lineshape $|E(\omega)|^2$ of the resonant mode is obtained by a fast Fourier transform of the time-trace. The $\lambda_{res}$ and $Q$ values are obtained by fitting the lineshape to a Lorentzian function:

\begin{equation}
  |E(\omega)|^2 \propto \frac{1}{1 + 4Q^2 (\frac{\omega}{\omega_{res}}-1)^2}
  \label{eqn:lineshape}
\end{equation}

\noindent where $\omega_{res} = 2\pi c / \lambda_{res}$ is the resonant frequency.

\indent Figure \ref{fig:figure_supp_optimization}(b) shows the wavelength and quality factor of the resonant mode as a function of $D$, for $L_1$ = 330 nm and $L_2$ = 120 nm. 
The grey dashed line indicates the diameter $D_W$ for which $\lambda_{res}$ = 1218 nm. 
The same series of calculations was performed for several $L_1$ and $L_2$ values. 
Figure \ref{fig:figure_supp_optimization}(c) displays $D_W$ as a function of $L_1$, together with the associated quality factor $Q_W$. 
A maximum quality factor of 167 is obtained for $D_W$ = 967 nm, $L_1$ = 330 nm and $L_2$ = 120 nm.

\indent The Purcell factor $F$, defined as the ratio between the emission rate of the dipole in the cavity and in bulk silicon, is obtained by a calculation of the steady-state electric field in the cavity. 
At time $t=0$, the dipole starts to emit continuous quasi-monochromatic radiation at 1218 nm, using a slow intensity ramp to avoid step-induced harmonics. 
The total emitted power is recorded by a monitor tightly localized around the dipole. 
The emitted power increases as the back-reflected electric field stimulates the emission of the dipole, until a stationnary regime is reached. 
The Purcell factor is given by the normalized power value in the stationnary regime. 
Figure \ref{fig:figure_supp_optimization}(c) shows the Purcell factor for each ($D_W$, $L_1$) set obtained from ringdown calculation. 
A Purcell factor of 13 is obtained for $D_W$ = 967 nm, $L_1$ = 330 nm and $L_2$ = 120 nm.

\newpage

\subsection{Effect of spectral detuning}

\begin{figure}[ht]
  \centering
  \includegraphics[width=240pt]{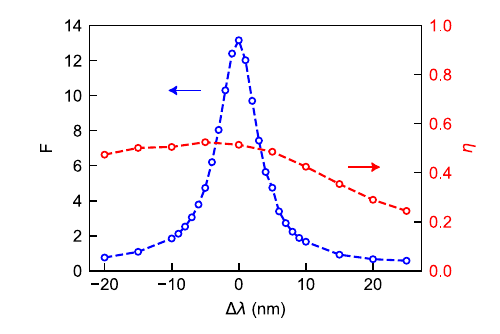}
  \caption{\textbf{Purcell factor and collection efficiency of the CBG cavity as a function of spectral detuning.} The spectral detuning $\Delta \lambda$ is the difference between the emission wavelength of the dipole and $\lambda_{res}$ = 1218 nm. For estimating the collection efficiency, a monitor records the intensity radiated through a square aperture located in the farfield region above the dipole. The solid angle for this aperture is the same as for a microscope objective with a numerical aperture of 0.65. The collection efficiency $\eta$ corresponds to the ratio between the intensity recorded by this monitor and the total intensity emitted by the dipole. A collection efficiency of 51$\%$ is obtained at zero-detuning, and remains above 40$\%$ in the [-20 - +10] nm detuning range.}
  \label{fig:figure_supp_cbg_vs_spectral_detuning}
\end{figure}

\newpage

\section{Quantum efficiency and zero-phonon emission rate}

\begin{figure}[ht]
  \centering
  \includegraphics[width=120pt]{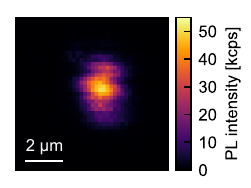}
  \caption{\textbf{PL map of the CBG-coupled W center at saturation under pulsed excitation.} 
  The excitation laser has a repetition rate of 5 MHz and average power 17.5 $\mu$W.
  A maximum PL intensity around 52 kc/s is observed when the pump beam is located at the center of the cavity.}
  \label{fig:figure_supp_scan_pulsed}
\end{figure}

\indent \textbf{Quantum efficiency.} The quantum efficiency of the W center under study is estimated from the PL intensity at saturation in the 1200-1300 nm spectral window under above-gap pulsed excitation, by a detailed analysis of the photon budget from generation to detection. More precisely, the aim is to estimate the probability $\eta_{ZPL}$ for the W center to relax through zero-phonon emission rather than other channels (non-radiative recombination, and phonon-assisted radiative recombination). Since the measured Debye-Waller is near unity, we can safely neglect the contribution of phonon-assisted emission. The PL intensity is thus proportional to $\eta_{ZPL}$ with good approximation:

\begin{equation}
    \label{eqn:count_rate_pulsed}
    \Gamma_d^{(pulsed)} = \eta_{ZPL} \times \eta_{coll} \times T_{fl} \times T_{setup} \times \\
    \eta_{SDE} \times \frac{\tau_{on}}{\tau_{on}+\tau_{off}} \times \Gamma_{exc}
\end{equation}

\noindent where:
\begin{itemize}
    \item $\Gamma_d^{(pulsed)}=52$ kcps is the PL intensity at saturation (Figure \ref{fig:figure_supp_scan_pulsed}).

    \item $\eta_{coll} = 51 \%$ is the collection efficiency of the PL by the microscope objective with a numerical aperture of 0.65, estimated by FDTD simulation.

    \item $T_{fl} = 72 \%$ is the sample-to-first lens transmission, taking into account the cryostat's window and the microscope objective. The estimation is based on the transmission values provided by the manufacturers.

    \item $T_{setup} < 56 \%$ is the transmission from the microscope objective to the photodetector, corresponding to the product of all the optics transmission values and fiber-coupling efficiencies. The estimation is an upper bound as it assumes perfect overlap between the PL mode and the fundamental mode of the collection fiber.

    \item $\eta_{SDE} = 81 \%$ is the detection efficiency of the SNSPD, as provided by the manufacturer.
    
    \item $\tau_{on} = 153 \pm 3$ ns and $\tau_{off} = 741 \pm 5$ ns are the characteristic times of the photoblinking behaviour, which must be taken into account in order to estimate the quantum efficiency when the W center is in its optically active state (i.e., $17.1 \pm 0.4 \%$ of the time).

    \item $\Gamma_{exc}=5$ MHz is the excitation laser repetition rate.
\end{itemize}

\indent Using Equation \ref{eqn:count_rate_pulsed}, we obtain $\eta_{ZPL} > 38 \%$. Several important comments can be made about this value:
(1) It is a lower bound as it assumes that the best possible sample-to-detector transmission value is reached (perfect overlap between the PL mode and collection mode, aberration-free confocal microscope, no scattering by the optics, etc\dots).
(2) This value is the quantum efficiency when the emitter is optically active, corresponding to a first-lens efficiency of 14\%. Taking photoblinking into account, the effective first-lens efficiency of the source drops to 2.4\%.
(3) Importantly, Equation \ref{eqn:count_rate_pulsed} is valid under the assumption that the average delay between successive detection events is much larger than the 29 ns recovery time of the SNSPD. This is the case as the repetition period is 200 ns in our experiment.
(4) We have implicitly assumed that one photon at most is detected for each pumping cycle, despite the fact that $A_{0} = 0.51 \pm 0.01$ at saturation indicates that about two photons are emitted on average per pulse. This assumption is perfectly valid as the delay between the emission of the two photons ($\approx$ 7 ns) is much shorter than the recovery time of the detector.\\

\indent \textbf{Zero-phonon emission rate.} The zero-phonon emission rate of the W center under study, given by $\Gamma_{ZPL} = \eta_{ZPL} / \tau$ where $\tau$ = 7 ns is the lifetime of the excited state, is therefore larger than 1/18 ns$^{-1}$. This value should be compared with the zero-phonon emission of 1/134 ns$^{-1}$ estimated for W centers in bulk silicon \cite{lefaucher_purcell_2024}, demonstrating Purcell enhancement of the zero-phonon emission rate for the cavity-coupled emitter by a factor $F > 7.2$.\\

\indent \textbf{Analysis of the CW count rate.} It is also possible to estimate $\eta_{ZPL}$ from the PL intensity at saturation under continuous-wave pumping:

\begin{equation}
  \label{eqn:count_rate_cw}
  I_{sat}^{(CW)} = \eta_{ZPL} \times \eta_{coll} \times T_{fl} \times T_{setup} \times \\
  \eta_{SDE} \times \frac{\tau_{on}}{\tau_{on}+\tau_{off}} \times \frac{1}{\tau}
\end{equation}

\noindent From the count rate of $1.29 \pm 0.01$ Mcps at saturation, we get $\eta_{ZPL} > 33 \%$, close to the estimate obtained under pulsed pumping. Nevertheless, since the photoblinking times were measured only under pulsed excitation, the estimation in the pulsed regime is more accurate.

\newpage

\section{Scanning electron microscopy image of the cavity}

\begin{figure}[ht]
  \centering
  \includegraphics[width=267pt]{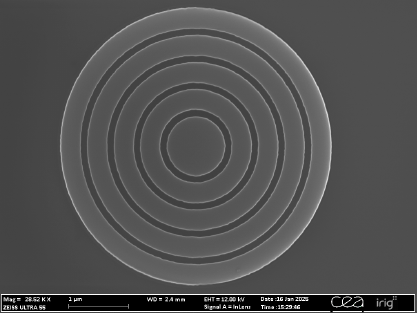}
  \caption{\textbf{Scanning electron microscopy image of a CBG cavity,} similar to the one containing the single W center studied in the manuscript.}
  \label{fig:figure_supp_meb_cavity}
\end{figure}

\newpage

\section{Photon autocorrelation at high power}

\begin{figure}[ht]
  \centering
  \includegraphics[width=267pt]{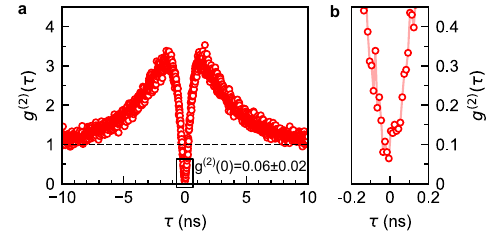}
  \caption{\textbf{Photon autocorrelation function of the W-1 center at high power.}
  \textbf{a,} Measurement is performed under above-gap CW excitation at $P = 9.2P_{sat}$.
  \textbf{b,} A zoom at short time delays reveals a $g^{(2)}(0)$ value of 0.06 $\pm$ 0.02}
  \label{fig:figure_g2_high_power}
\end{figure}

\newpage

\bibliography{bib_file}